\documentclass[aps,prl,groupedaddress,superscriptaddress,twocolumn,10pt,longbibliography]{revtex4-2}
\pdfoutput=1 
\usepackage[utf8]{inputenc}
\usepackage{hyperref}
\usepackage{graphicx}
\usepackage{orcidlink}
\usepackage{booktabs}
\usepackage{tabularx}
\usepackage{amssymb}
\usepackage{physics}
\usepackage{mathtools}
\usepackage{lipsum}
\newcommand{\out}[1]{}

\renewcommand{\vec}[1]{\mathbf{#1}}

\begin{document}

\title{
Universal response of Rydberg manifolds to standing light waves  from the microwave to the X-ray regime}

\author{Homar Rivera-Rodr\'iguez \orcidlink{0009-0007-6454-7063}}
\affiliation{Max-Planck Institut für Physik komplexer Systeme, Nöthnitzer Stra\ss e 38, 01187 Dresden, Germany}

\author{Matthew T. Eiles\orcidlink{0000-0002-0569-7551}}
\affiliation{Max-Planck Institut für Physik komplexer Systeme, Nöthnitzer Stra\ss e 38, 01187 Dresden, Germany}
\affiliation{Department of Physics and Astronomy, Purdue University, West Lafayette, IN 47907, USA}

\author{Jan M. Rost}
\affiliation{Max-Planck Institut für Physik komplexer Systeme, Nöthnitzer Stra\ss e 38, 01187 Dresden, Germany}

\date{\today}

\begin{abstract}
\noindent
Standing light waves structure the electronic density of a Rydberg atom in a rich 
 but surprisingly systematic fashion.
 We uncover these systematics, which are nearly universal across a large range of principal quantum numbers $n$, by varying the wavelength of the standing light over several orders of magnitude. Thereby, we identify five qualitatively different regimes and give their transition criteria in terms of specific critical wavelengths.  
 The bandwidth of the lattice spectrum, manifested in the difference of energies between the states on the edges of the degeneracy-lifted $n$-manifolds, as
 well as the organization of the electron density in coordinate and momentum space are used to rationalize the systematics. A experimental setup is proposed to measure the features in the different regimes.
\end{abstract}

\maketitle
Optical lattices have played a pivotal role in structuring ultracold gases into assemblies of atoms following a regular spatial pattern \cite{Bloch2005} typically experimentally realized at sub-micrometer-scale periods.
In a second, largely independent direction, the shaping of intense higher frequency fields  with harmonic techniques and X-ray free electron lasers is rapidly progressing \cite{jark99,shi25} with the potential to generate optical lattices  of much smaller spatial periods. Following a third thread, ultracold platforms now routinely produce very large Rydberg orbitals with principal quantum numbers well beyond $n=100$ \cite{Balewski2013, Schlagmuller2016}. Scaling as $n^2 a_0$, these orbitals reach extensions comparable to the spatial periodicity of optical lattices. 
For Rydberg atoms, the ponderomotive shift induced by a standing wave depends on the overlap between the intensity profile and the electronic wavefunction. The effect has been mainly exploited at wavelengths relevant to trapping and spectroscopy, where the atom size is typically small compared to the lattice period \cite{Duta2000,Younge_2010_adiabatic,Younge_2010, Younge2010_prl, Anderson2011, Anderson2012, Anderson2013, Topcu2013, Topcu2013_2, Moore2015, Wang_2016, Malinovsky2020, Cardman_2021, Cardman2023}.

Here, we bring these three threads from the ultrafast and ultracold domains together by systematically analyzing how a standing light wave modifies the spectrum of a Rydberg electron as the lattice wavelength is varied from the long-wavelength limit, associated with the well-known Stark effect, to extremely short wavelength, where the lattice acts as an increasingly localized probe of the electronic wavefunction near the ionic core. We will show that
in between these two extremes the Rydberg manifold qualitatively reorganizes four times in response to the applied light wave, giving rise to five different regimes. Remarkably, these regimes are to a large extent universal; that is, the electronic levels as a function of the scaled lattice wavenumber are nearly the same over a large range of $n$.

We focus on a one-dimensional lattice formed by two linearly polarized light fields of amplitude $F_0$ counter-propagating  in the $z-$direction with wavelength $\lambda {=} 2\pi/k$, i.e., frequency $\omega {=} c \, k$. Since the optical frequency is several orders of magnitude larger than the characteristic time-scales of the relative electron-core and center of mass motions, the electron-light interaction is treated adiabatically in the Coulomb gauge with an effective {\it time-independent} but {\it spatially dependent} ponderomotive potential $V_{\mathrm{P}}$ added to the electronic Hamiltonian~\cite{Duta2000}. The resulting ponderomotive force is responsible for well-known phenomena such as the high-intensity Kapitza-Dirac effect~\cite{Bucksbaum1988} which has recently regained considerable attention with  elegant experiments using pulsed standing light waves \cite{lise+24,guli+25}. The ponderomotive potential for the Rydberg electron acquires the form (atomic units are used unless stated otherwise) 
\begin{equation}
	V_{\mathrm{P}}(\boldsymbol{r};Z_0)=\frac{1}{2} V_0(1+\cos[2k(z+Z_0)])\,,
	\label{eq:Vp}
\end{equation}
where $V_0=F_0^2/ \omega^2$, the electron position is $\vec r$, and the lattice is shifted by a distance $Z_{0}$ relative to the ion.  We are interested in the principal structural effects the standing light field has on the Rydberg electron density and how these depend on $k$. Therefore, we set $Z_0{=}0$ and vary $k$ at fixed $V_0$ to isolate the role of the lattice spacing. 
In the same vein, to concentrate on the effect of $V_P$, we consider a hydrogenic electron and neglect fine structure so that all states with fixed principal quantum number $n$ are degenerate. Under the full Hamiltonian, $H = \vec p^2/2-1/r {+}V_{\mathrm{P}}(\boldsymbol{r};0)$, different $n-$manifolds do not mix provided $V_0$ is sufficiently small.  However, for a given $n$ the angular momentum $\ell$ is no longer conserved while $m$ remains a good quantum number since $V_{\mathrm{P}}$ respects cylindrical symmetry \footnote{For the choice of $Z_0{=}0$ the potential has reflection symmetry, $V_P(\vec r,0)=V_P(-\vec r,0)$, rendering the eigenenergies for $\pm m$ degenerate. We will use $m>0$ for both values $\pm m$.
}.

In fact, the energy for the state with maximal $m= n-1$ is given analytically by 
\begin{equation}\label{eq:circ}
  E^{(n-1)}   = \frac{V_0}2\left [1+\left(1+\frac{b^2}{2n}\right)^{-(n+1)}\right]\,,
\end{equation}
where $b$ is the scaled lattice wavenumber
\begin{equation}
 b=\sqrt{2} \, n^{3/2} \, k\,.
 \label{eq:scaled_b}
\end{equation}
Then,   we obtain  $E^{(n-1)}(b) \!\approx \! E(b)=\tfrac{V_0}{2}(1+e^{-b^2/2})$ for large $n$,
such that the only dependence on $n$ comes from Eq.~\eqref{eq:scaled_b}.
In general, the full Hamiltonian $H$ must be diagonalized  numerically among the hydrogenic states with fixed $(n,m)$  which yields $n-m$  energy levels. We will concentrate in the following on the largest submanifold $m=0$, and in particular on its ``edge'' states having lowest ($E_{\rm low}$) and highest ($E_{\rm high}$) energy. The edge energies are shown    in Fig.~\ref{fig:delta} for $n=30$, $60$, and $90$ alongside 
with  their difference normalized to $V_0$,
\begin{equation}
\Delta(b)=\left [E_{\mathrm{high}}(b)-E_{\mathrm{low}}(b)\right] /V_0\,.
\label{eq:splitting}
\end{equation}

\begin{figure}[t]
	\centering
    \centering
	\includegraphics[width=\linewidth]{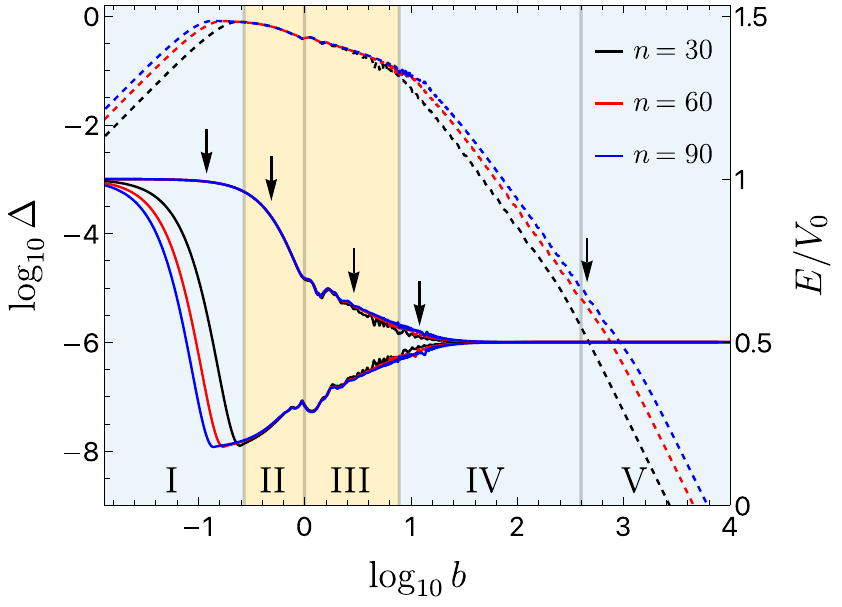}
	\label{fig:delta}

        \centering
        \begin{tabular}{c|c|c|c|c|}
		 Transition \hfill & \hfill I$\to$II \hfill & \hfill II$\to$III \hfill & \hfill III $\to$ IV  \hfill & \hfill IV $\to$ V \hfill \\
		\hline
		$b_{A\to B}$ & $\frac{\pi}{3} \sqrt{\frac{2}{n}}$ & 1 & $\sqrt{2n}$ & $\frac{\pi}{\sqrt{2}} n^{3/2}$ 
	\end{tabular}

\caption{Energy splitting $\Delta(b)$ (dashed) and edge energies $E_{\rm high}(b)$, $E_{\rm low}(b)$ (solid) of the $m=0$ manifold for  $n=30$, $60$, and $90$. The vertical gray lines demark the five regimes for $n=30$ and the arrows denote the representative $b$ values of the densities shown in Fig.~\ref{fig:densities2d}. The table gives $b_{\rm A\to B}$ at the transitions, see text.}
\label{fig:delta}
\end{figure}

In the limits $b{\to} 0$ and $b{\to} \infty$, the ponderomotive potential produces only a constant shift and all states (including all $m$ subspaces) with the same $n$ are degenerate. {Therefore, in these limits all states have the same energies as  the state with maximal $m$, Eq.~\eqref{eq:circ}, namely $\lim_{b \to 0}V_{\rm P}(\vec r)\to V_0$ in the long-wavelength 
(DC) limit and $\lim_{b \to \infty}V_{\rm P}(\vec r)\to V_0/2$ as the high-frequency oscillations in the standing light wave average out over the electron density. 
For finite $b$, the lattice lifts the  degeneracy. 
Looking at the edge energies in Fig.~\ref{fig:delta}, we see that $E_{\rm high}/V_0$ shows essentially a monotonic decrease from $1$  to $1/2$ following the behavior of the state with maximal $m$ apart from local irregularities. The scaling with $n$ is almost as perfect as for the maximal $m$ state. Since the electronic state with maximal $m$ has nodes only in the direction transverse to the lattice, it lacks the flexibility to minimize the energy with respect to the ponderomotive potential's variation along the $z$ axis. On the other hand, $E_{\rm low}$ shows the largest energetic variation along $b$ with a pronounced minimum, since here the wavefunction nodes are distributed dominantly along $z$, as we will see.

The vertical lines in Fig.~\ref{fig:delta} separate the five regimes in which the states behave differently, most clearly visible in the scaling behavior and slopes in the difference of the edge energies, $\Delta(b)$, shown in Fig.~\ref{fig:delta} as dashed lines. In the intermediate regimes II and III the scaling through $b$ with $n^{3/2}$ is almost perfect,
while the scaling in the regimes I, IV, and V is clearly different, setting the regimes apart from each other. However, do these regimes also feature in measurable observables?

\begin{figure}[t]
	\centering
	\includegraphics[width=\linewidth]{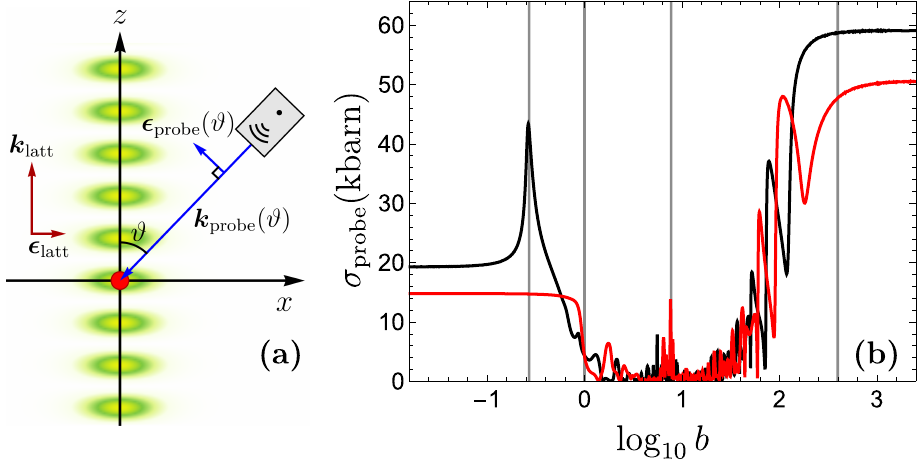}
	\caption{(a) Probe photoionization geometry, where $\vartheta$ is the angle between the probe field and lattice field propagation vectors and the Rydberg core (red circle) is located at the origin. (b) Photoionization cross section $\sigma_{\mathrm{probe}}(b)$ for a perpendicular probe field 
    ($\vartheta = \pi/2$) with $\lambda_{\rm probe}=1064$ nm for the  edge states with $E_{\rm low}$ (black) and $E_{\rm high}$ (red) in the $n=30$ manifold for lattice depth $V_0=2$ GHz. The gray vertical lines mark the transitions between regimes as in Fig.~\ref{fig:delta}.}
	\label{fig:pi_probe}
\end{figure}

 Indeed, the photoionization cross section of the edge states by a weak probe field shown in Fig.~\ref{fig:pi_probe} reveals considerable differences when $b$ is varied, with a tendency towards small cross sections for intermediate $b$ and larger ones in the limits of small and large $b$. The five different regimes (marked with thin vertical lines) can be identified from features in the cross section. The first region of large cross sections for small $b$ is  subdivided into regimes I and II by a characteristic peak in $\sigma_{\rm low}$. A smaller but pertinent peak in the cross section $\sigma_{\rm high}$ separates regimes III and IV in the region of intermediate $b$, while the region of large $b$ is characterized by  large cross sections in the single regime V. Note, that these features do not sensitively 
 depend on the energy of the ionizing photon.

 \begin{figure*}[tp]
\centering
\includegraphics[width=\linewidth]{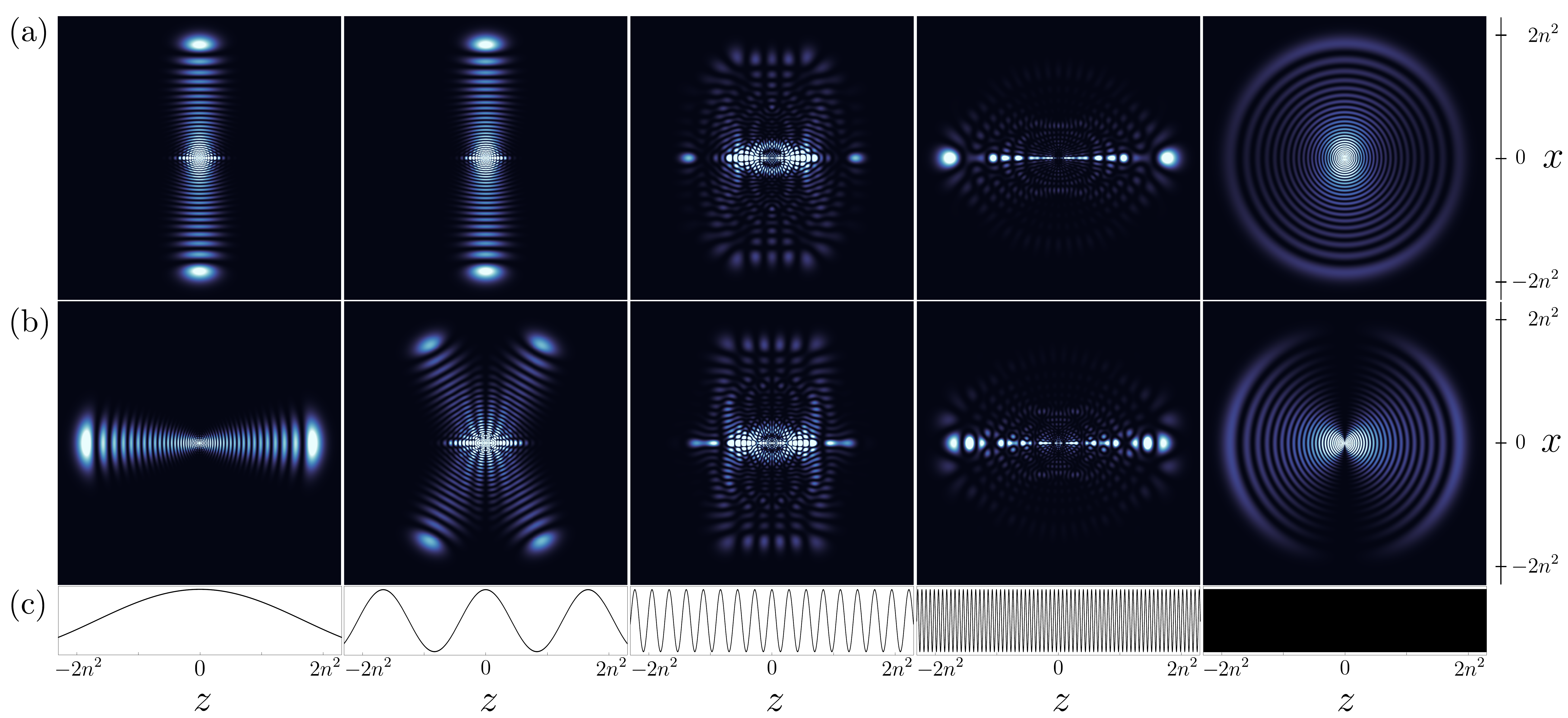}
\caption{Electron density $\rho_\mathrm{Ryd}\equiv|\psi(\boldsymbol{r)}|^2$ in the $xz$ plane for the upper (a) and lower (b)  edge states for $m=0$ and $n=30$ at the representative values $ \log_{10}b=-0.9, \ -0.3, \ 0.5, \ 1.1, \ 2.7$ marked by vertical arrows in Fig.~\ref{fig:delta}. Panel (c) shows the  ponderomotive potential of the optical lattice at the corresponding wavelengths. For reference the box size is approximately $4n^2 a_0=190$ nm.} 
  \label{fig:densities2d}
\end{figure*}

 In order to understand the systematics behind the different regimes, in Fig.~\ref{fig:densities2d} we take a look at representative electron densities of the edge states in each regime alongside the respective potentials. The guiding organizational principle of the system is the adaptation of the Rydberg electron density $\rho_\mathrm{Ryd}$ built from the Hilbert space of a fixed $(n,m)$ manifold to the shrinking spatial period $\lambda/2 = \pi/k \propto 1/b$ of the ponderomotive potential as $b$ increases. For the lower edge state, $\rho_\mathrm{Ryd}$ is localized as much as possible in the valleys of the potential. There is no valley in regime I since the period of the potential is much larger than the maximal extension of $\rho_\mathrm{Ryd}$, $\lambda\gg n^2$. Rather, the potential can be approximated by two linear branches, symmetric about $z=0$.
Hence, the maximal extension of $\rho_\mathrm{Ryd}$ along $z$ is provided by the extreme parabolic states with $\rho_\mathrm{Ryd}$ concentrated near $z_n=\pm 3n^2/2$ (see End Matter). This changes in II where $\lambda/4 \lesssim |z_n|$ 
 and the foci of $\rho_\mathrm{Ryd}$ move closer together following the location of the valley minima. Consequently, $\lambda/4 = 3n^2/2$ marks the transition with the corresponding $b_{\rm I\to \rm II} = (\pi/3) (2/n)^{1/2}$ shown as a vertical line
 in Figs.~\ref{fig:delta} and~\ref{fig:pi_probe}. Note that the higher edge state does not change qualitatively in I and II as it remains localized on the central maximum of the potential at $z{=}0$ with an essentially Gaussian width $\sigma_n=n^{3/2}$  (see End Matter).  However, the upper edge state affects the transition from II to III. This transition
 represents the switch from spatially dominated structures to momentum dominated structures at $b_{\rm II\to \rm III}=1$, where the width $\sigma_n$ matches the Gaussian approximated width of the lattice extrema.

 Once several optical wavelengths fit within the Rydberg orbital, the highest and lowest edge energies are no longer controlled by the isolated location of $\rho_\mathrm{Ryd}$ in real space.  Instead, the relevant property is the longitudinal momentum coherence induced by the lattice, i.e., how the multinodal density pattern of the Rydberg atom matches, or can be approximately commensurate with, the lattice structure. This is a consequence of the ponderomotive term $\cos(2kz)$ which couples momentum components $p_z$ separated by $2k$.   After tracing over the transverse momenta to obtain the reduced  momentum density matrix $\varrho(p_z,p_z')$, the ponderomotive shift reads $E(k)\propto \tfrac{1}{2}+\tfrac{1}{2} \Re \left( \!\int \varrho(p_z,p_z+2k) \, dp_z \right)$. Hence, in the momentum-dominated regimes, $\Delta(b)$ is determined by the off-diagonal elements of the reduced momentum density matrix, specifically those which connect electron momenta differing by the lattice momentum. Now the difference
 of the coherent $\rho_\mathrm{Ryd}$ of the two edge states is a consistent phase difference of $\pi$, localizing $\rho_\mathrm{Ryd}$ largely on the maxima or minima  of the ponderomotive potential for $E_{\rm high}$ or $E_{\rm low}$, respectively.
 
In regime III there are sufficiently many nodes available in the $(n,0)$ manifold to establish coherence with the ponderomotive potential over all wavelengths contained within the classically allowed region of the electron. This is possible until 
 $n k=1$, when the Rydberg volume contains $\sim n$ half-wavelengths, comparable to the $\sim n$ nodes available to the wave function. This implies
 $b_{\rm III\to IV} \!= \!\sqrt{2n}$ as the transition between regimes III and IV.  In the density matrix perspective for $b>b_{\rm III\to\rm IV}$, the momentum separation $2k$ imposed by the lattice exceeds the momentum range over which the Rydberg state has appreciable weight, causing the shifted off-diagonal density matrix elements to  decrease for increasing $b$.

In regime V the ponderomotive lattice oscillations are too rapid to be resolved by the Rydberg electron density everywhere except close to the ionic core as $\vec r \to 0$.
Consequently, the upper and lower edge states evolve toward $s$- and $p$- states respectively, slightly asymmetric about $E=V_0/2$.  This happens if the lattice period  $\lambda/2 =\sqrt{2} \,\pi \, n^{3/2}/b$ is smaller than the distance  $r^* \approx 2$ {(see End Matter)} to the first node in $\rho_\mathrm{Ryd}$ leading to  $b_{\rm IV\to V}= \pi\,  n^{3/2}/ \sqrt{2}$.

The five  regimes we have identified  persist in multidimensional ponderomotive lattices as illustrated in Fig.~\ref{fig:spectra_multi}, which shows the Rydberg spectrum in the presence of one-, two-, and three-dimensional lattices formed from mutually perpendicular standing waves (see End Matter). Although $m$ is no longer conserved in the higher-dimensional cases, the five regimes remain visible in the splitting $\Delta(b)$, now defined from the edge states of the full $n-$manifold. Increasing the lattice dimensionality compresses the spectrum, leading to a smaller spread in  $\Delta(b)$, and shifts the position of the crossover between regimes I and II. The shift has a simple geometric origin: distributing the electron density over $D$ lattice directions reduces the typical extent per dimension by roughly $1/\sqrt{D}$, giving $b_{\rm I\to II}^{(D)}\simeq \sqrt{D}\, b_{\rm I\to II}$. The remaining crossovers are controlled by the axial width, momentum scale and near-core structure identified above, all of which are only weakly affected by the lattice dimensionality. In fact, the transition from IV to V is most clearly visible from this perspective,
since the point-like perturbation through the ponderomotive potential for $b\to \infty$ is not sensitive to the dimensionality of the potential and therefore all $\Delta(b)$ merge to a single curve at $b_{\rm IV\to \rm V}$ as given above.

\begin{figure}[t]
	\centering
	\includegraphics[width=\linewidth]{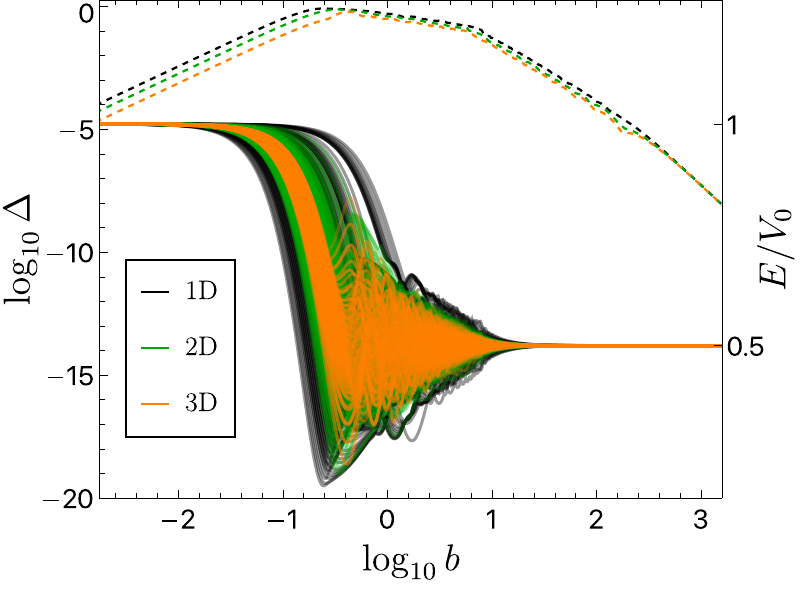}
	\caption{Spectrum (solid) and (b) edge splitting $\Delta(b)$ (dashed, Eq.~\ref{eq:splitting}) for $n=30$ in 1D, 2D, and 3D lattices. The intensity for each standing light wave is scaled to match together the intensity in the 1D case. The 1D case shown here includes all  $m-$submanifolds.}
	\label{fig:spectra_multi}
\end{figure}
We conclude with a discussion of the feasibility of accessing the regimes and their phenomena experimentally. 
Direct Rydberg electron photoionization by the lattice field is negligible with respect to the ponderomotive effect in the parameter range considered here. We quantify it by the dimensionless ratio $\chi=\Gamma_{\rm PI}/V_0=\sigma_{\rm PI}/(2\pi\alpha/\omega)$, where $\Gamma_{\rm PI}$ and $\sigma_{\rm PI}$ are the lattice-induced photoionization level width and cross section respectively, and $\alpha$ is the fine-structure constant. For $30\le n\le60$ and wavelengths down to $10\,{\rm nm}$, we find $\chi\lesssim10^{-3}$.
Since $\chi$ scales empirically as $n^{-3}$ for fixed wavelength \cite{Younge_2010_adiabatic}, this effect is even weaker at larger $n$.  The more restrictive short-wavelength limitation is core photoionization. For Rb, the outermost core-shell ionization threshold is 27.28 eV, corresponding to $\lambda \simeq$ 45 nm \cite{NIST2024}; avoiding this processes therefore requires $\lambda \gtrsim45$ nm. Even with this constraint, the III-IV crossover can be reached at sufficiently large $n \! \sim \! 150$, while lighter alkali atoms relax this bound further. For Li and Na the corresponding thresholds allow wavelengths down to roughly 16 nm and 26 nm, respectively~\cite{NIST2024}, extending experimental access to regime IV at more moderate $n \! \sim \! 90$. Pushing deeper into regime IV would require still shorter wavelengths in the soft–X–ray domain, where the induced splittings are already much smaller and the regime becomes less attractive spectroscopically.

From an experimental perspective, observing by whatever means all five regimes at a fixed $n$ is not feasible, since $b$ (and consequently $\lambda$) spans several orders of magnitude requiring very different experimental setups. It is more realistic to choose $n$ and the lattice wavelength $\lambda$ to determine $b$ in the desired regime. For typical ponderomotive potentials with $V_0 \approx 2$ GHz, considering an isolated manifold $n$ remains well justified since the induced splittings remain smaller than the separation between neighboring $n-$manifolds up to $n \sim 130$, with this upper limit increasing for shallower lattices.
Once the lattice-dressed states have been created,
photoionization provides a sensible and sensitive probe since the photoionization cross section $\sigma_{\mathrm{probe}}$ varies strongly and systematically with $b$.
This robustly uncovers the five regimes even under different probe wavelengths, as shown in Fig.~\ref{fig:pi_probe}(b). However, the insight, again, comes only from a synopsis over a wide range of lattice wavenumbers $b$.
\begin{figure}[t]
	\centering
	\includegraphics[width=\linewidth]{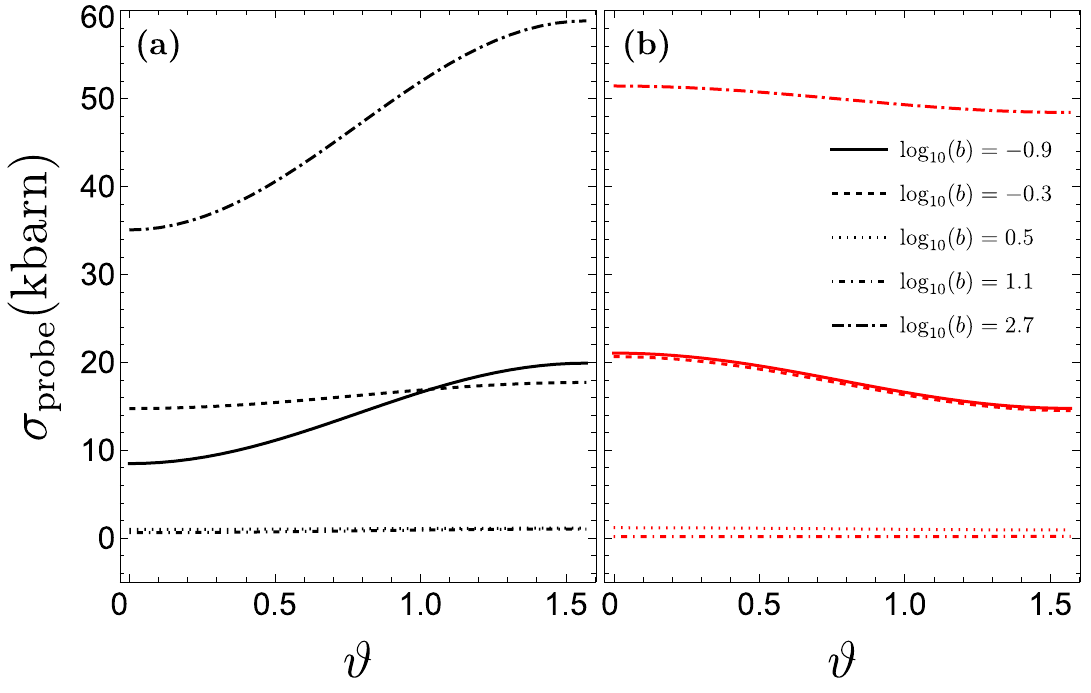}
	\caption{Probe photoionization cross section as a function of the probe-lattice angle $\vartheta$ for the lower (a) and upper (b) edge $m=0$ states at the representative lattice wavenumbers $b$ corresponding to the wavefunctions of Fig.~\ref{fig:densities2d}.}
	\label{fig:sigma_theta}
\end{figure}

A complementary, experimentally feasible protocol is to rotate the probe geometry at fixed $b$.  We predict in Fig.~\ref{fig:sigma_theta} a distinct angular cross section $\sigma_{\rm probe}(\vartheta)$ for each $b$, allowing the regime-dependent state character to be read out without changing the lattice wavelength. In brief, $\sigma_{\rm probe}(\vartheta)$ changes oppositely for lower, Fig.~\ref{fig:sigma_theta}(a), and upper, Fig.~\ref{fig:sigma_theta}(b), edge states, and the cross section is almost independent of $\vartheta$ in the momentum dominated regimes III and IV.

To summarize, we have analyzed the response of a Rydberg electron to a standing light wave potential with a wavenumber varying over several orders of magnitude. Five different regimes emerge characterized by distinctly structured electron densities,
which exhibit distinctive responses when probed via photoionization. 
The character of these states bears some similarity to those predicted to occur in Rydberg composites by exciting a Rydberg atom embedded in a regular array of ground-state atoms such that the electron interacts with an array of delta-function potentials \cite{Hunter2020}.  What makes optical engineered obstructions special is the control of the width of the perturbing periodic potential to structure Rydberg electron densities as demonstrated here. This property of light-induced forces has also been used in a recent  optical tweezer proposal \cite{Rivera2026}, where light focused to a waist smaller that the Rydberg orbit generates sub-orbital ponderomotive forces that enable eccentric trapping and sculpt localized Rydberg electron wavefunctions.
We expect the flexible control over widths and positions of pondermotive potentials 
to open ample ways to prepare and stabilize fragile electronic states in the future.

\bibliography{references.bib}  
\clearpage
\makeatletter 
\renewcommand{\thefigure}{S\@arabic\c@figure}
\makeatother

\makeatletter 
\renewcommand{\theequation}{S\@arabic\c@equation}
\makeatother

\setcounter{equation}{0}
\setcounter{figure}{0}

\section*{End matter}

\subsection{A. Reduced Rydberg densities and Gaussian representation}

The reduced axial density of the Rydberg electron is
\begin{equation}
\rho_z(z)=\int_{-\infty}^{\infty} \int_{-\infty}^{\infty} \left| \Psi(\vec{r}) \right|^2 dx \, dy.
\label{eq:rho_def}
\end{equation}
Figure~\ref{fig:densities} compares $\rho_z(z)$ of the edge $m=0$ states with the ponderomotive profile $V_{\rm P}(z)$ for the representative wavelengths of Fig.~\ref{fig:densities2d}. In regimes I and II the spatial relations are simple: the high-energy state is localized near the lattice maximum, while the low-energy state avoids it and, in regime II, localizes near the central lattice minima. This direct density-lattice correspondence underlines the Gaussian description. 
In regimes I and II $\rho_z(z)$ can be accurately represented using $\rho_z(z)=G^2(z;\sigma_n, z_n)$ where
\begin{equation}
G(z; \sigma_n, z_n)=\frac{1}{(2\pi \sigma^2_n )^{1/4}} \exp \left[-\frac{(z{-}z_n)^{2}}{4 \sigma^{2}_n} \right],
\end{equation}
is a Gaussian profile with normalization $1=\int\rho_z(z)\mathrm{d}z$. 
Here $z_n$ is the Gaussian distribution's peak position and $\sigma_n^2$ its variance. 
The upper edge state in regimes I and II is centered at the lattice maximum, so $z_n \!= \!0$. In the parabolic basis, this state is a narrow packet of balanced $n_1,n_2$ components centered near $n_1{-}n_2 \!= \!0$. Its expansion has a Gaussian envelope with width $\sqrt{n}$ that implies $\sigma_n= n^2/ \sqrt{n} =n^{3/2}$. 

\begin{figure}[h]
	\centering
	\includegraphics[width=\linewidth]{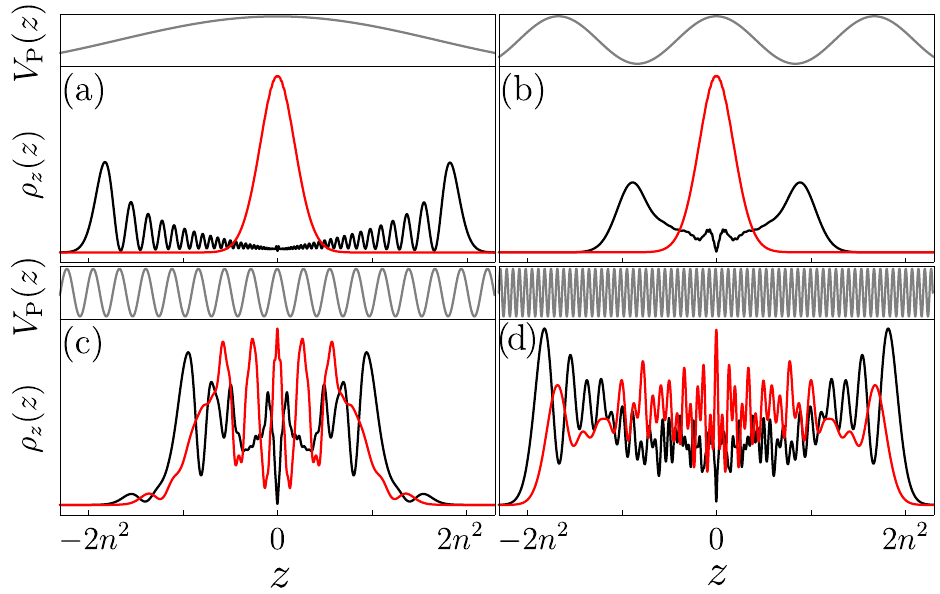}
	\caption{Reduced axial densities $\rho_z(z)$ for the lower (black) and upper (red) edge states for $n=30$, shown together with the ponderomotive potential $V_{\rm P}(z)$. Panels (a)-(d) correspond to the same scaled wavenumbers as in Fig.~\ref{fig:densities2d}: $ \log_{10}b=-0.9, \ -0.3, \ 0.5$ and $1.1$ respectively.}
	\label{fig:densities}
\end{figure}

The regime I low-energy state is the antisymmetric superposition of the two stretched parabolic states $(|n_1,0,0\rangle {-} |0,n_2,0\rangle)/\sqrt{2}$ with $n_1=n_2=n-1$. Each stretched component has $\langle z \rangle =\pm \tfrac{3}{2}n(n-1)\approx \pm \tfrac{3}{2}n^2$, giving $z_n=\tfrac{3}{2} n^2$. Approximating this linear combination by two displaced Gaussians $
\psi_{-}(z) \propto G(z;\sigma_n, z_n) - G(z;\sigma_n, -z_n)$, and neglecting the small overlap between the two packets gives $\langle z^2\rangle=z_n^2+\sigma_n^2$. As $\langle z \rangle=0$ for this antisymmetric Gaussian combination the variance is simply $\langle z^2\rangle$. Matching this to the exact stretched-pair variance, $\sigma^2_{\mathrm{para}}=\frac{n^2}{2}\left(5n^2-9n+6\right)
$, with $z_n=\tfrac{3}{2}n^2$ yields $\sigma^2_n=\frac{n^4}{4}-\frac{9n^3}{2}+\frac{3n^2}{n}\simeq\frac{n^4}{4}-\frac{9n^3}{2}$. Thus the lower edge state has axial extent $\sigma_n^{(\rm low \ I)} \sim n^2$ rather than the $n^{3/2}$ of $\sigma_n^{(\rm high \ I)}$  explaining the imperfect collapse of the low-energy regime-I curves across different $n$ in Fig.~\ref{fig:delta}.

The I$\to$II crossover occurs when the wavelength is small enough that the first two lattice minima, located at $z_{\rm min}=\pm \tfrac{\pi}{2k}=\pm \tfrac{\pi n^{3/2}}{\sqrt{2} b}$, enter the region accessible to the electron in the lower edge state given by $z_n^{(\rm low \ I)}=\frac{3}{2}n^2$. Then, imposing $z_n^{(\rm low \ I)}=z_{\rm min}$ gives the crossover presented in the main text.

In regime II, the low-energy density follows the central lattice minima, $z_{\rm min}$, and the corresponding center is $z_n=\pi n^{3/2}/(\sqrt{2}\, b)$. The numerical density retains a small central weight which is not captured by the approximate double Gaussian form.
To emulate this we use an effective inward-shifted center $z_n= \kappa \pi n^{3/2}/(\sqrt{2}\, b)$ with $\kappa <1$ determined numerically.

The II$\to$III crossover is estimated by matching the axial width of the upper edge reduced wavefunction to the effective lattice width. As established above, for the upper edge state the approximation $\rho_z(z)=G^2(z; \sigma_n,0)$ with $\sigma_n=n^{3/2}$ is quite accurate. Using this to approximate the axial wavefunction $\psi_z(z)=\sqrt{\rho_z(z)}$, the Fourier transform $\tilde{G}(p_z) \propto e^{-\sigma_n^2 p_z^2}$ can be used to obtain the oscillatory part of the ponderomotive shift as 
\begin{equation}
\langle \cos(2kz) \rangle=\int_{-\infty}^{\infty} \tilde{G}(p_z) \tilde{G}(p_z-2k) dp_z=e^{-2k^2 \sigma_n^2},
\end{equation}
where the waist used here is $\sigma_n=\sigma_n^{(\rm high \ I)}=n^{3/2}$. This exponential can be written in terms of an effective lattice width. Since, in regime II, $k\lesssim 1/n$, we assign a local Gaussian width to the lattice by expanding $\frac{V_{\rm P}(z)}{V_0}= \cos^2(k z)\approx 1-k^2 z^2$, expanding the Gaussian  $e^{-z^2/(2\sigma_{\rm latt}^2)}=1-\frac{z^2}{2\sigma_{\rm latt}^2}+\ldots$, and matching the quadratic terms to obtain the effective width $\sigma_{\rm latt}=\frac{1}{\sqrt{2}k}$. Therefore  $e^{-2k^2 \sigma_n^2}=e^{-(\sigma_n/\sigma_{\rm latt})^2}$. We estimate the crossover to occur when the widths coincide: $\sigma_n/\sigma_{\rm latt}=1\implies  b_{\rm II\to III} =1$.

Beyond estimating the crossovers, the Gaussian representation also provides simple approximate analytical expressions for the edge energies in the regimes where this representation remains valid. For any operator depending only on $z$, the expectation value is determined entirely by $\rho_z(z)$. In particular the ponderomotive energy at $Z_c=0$ is 
\begin{equation}
 E(b){=}\int_{-\infty}^{\infty} \rho_z(z) \, V_{\text{P}}(z;0) \,dz.
\end{equation}
For the Gaussian density $G^2(z;\sigma_n,z_n)$ this integral can be evaluated analytically giving  
\begin{equation}
E(b)=\frac{V_0}{2} \left[1{+} \exp\left( -\frac{\sigma_n^2 b^2}{n^3} \right) \cos(\sqrt{2 } \, z_n \,b/n^{3/2}) \right].
\label{eq:egauss}
\end{equation}

Using Eq.~\eqref{eq:egauss} with the values of $z_n$ and $\sigma_n$ obtained above provides a good approximation to the numerically computed edge energies in regimes I and I. For the lower edge state in regime II, in order to get a better agreement we use $\kappa \simeq0.72$, independent of $n$. As discussed in the text, this approximately accommodates the additional non-Gaussian accumulation 
in the electronic density near $z=0$. 

\section{Large $k$ asymptotics: regime V}
In regime V, the high- and low-energy states are the spherical $ns$ and $np$ ($m=0$ states) respectively. Up to sixth order of inverse powers of $k$, the asymptotic expectation values of the oscillatory part of the perturbation for $k \gg1$ are
\begin{align}
\big\langle ns,\,m{=}0 \big| \cos(2k z) \big| ns,\,m{=}0 \big\rangle&=
\frac{4n^{3}}{b^{4}}-\frac{4n^{4}(7n^{2}+5)}{3b^{6}}\\
\big\langle np,\,m{=}0 \big| \cos(2k z) \big| np,\,m{=}0 \big\rangle&=
-\frac{20n^{4}(n^{2}-1)}{3b^{6}}.
\end{align}
The crossover into this regime is estimated by comparing the lattice period $\lambda/2$ with the distance to the first radial node of $\rho_\mathrm{Ryd}$. If $r^*$ denotes the distance from the origin to this first density node, we define the crossover by $\lambda_{\rm IV\to V}/2=r^* $. Since $k=2 \pi/ \lambda$, this gives $b_{\rm IV\to V}= \frac{\pi}{r^*}\, \sqrt{2} n^{3/2}$. For a pure hydrogenic state, the first node of $\rho_\mathrm{Ryd}$ coincides with the first radial node of the $ns$ wavefunction. In the large $n$ limit this node approaches $j_{1,1}^2/8 \approx 2$  where $j_{1,1}$ is the first zero of the Bessel function $J_1(x)$. 

\section{2D and 3D lattices}
For the one-dimensional optical lattice described in the main text, the electric field is
\begin{align}
	\boldsymbol{F}(\boldsymbol{r},t)&=F_0 \, ( \cos(kz-\omega t)+\cos(-kz-\omega t)) \boldsymbol{e}_x \nonumber \\
	&=2F_0 \cos(\omega t) \cos (k z) \boldsymbol{e}_x.
\end{align}
Consider now 2D and 3D optical lattices. The electric field for the 2D lattice propagates along the $x$ and $y$ directions, and for the 3D lattice along the direction of all three coordinate axes. For identical laser settings ($F_0, k$) in each direction and the core located at $\boldsymbol{R}_0=\boldsymbol{0}$ the ponderomotive potentials are
\begin{equation}
V_{\mathrm{P}}^{(\mathrm{2D})}(\boldsymbol{r})=\frac{1}{2} V_0^{(\mathrm{2D})} \big[2+\cos(2 k x) +\cos(2ky) \big]
\end{equation}
and
\begin{align}
	V_{\mathrm{P}}^{(\mathrm{3D})}(\boldsymbol{r})=\frac{1}{2} V_0^{(\mathrm{3D})} \left[ \right. 3+\cos(2 k x) + \cos(2ky)+\left. \cos(2kz) \right].
\end{align}

To compare the results for different lattice dimensions at fixed total field intensity, the $D$-dimensional depth is scaled as 
\begin{equation}
V_0^{(\mathrm{D})}=\frac{1}{D} V_0^{(\mathrm{1D})}, 
\end{equation}

 In this cases the ponderomotive force acts along multiple directions but we set $\boldsymbol{R}_{0}=\boldsymbol 0$; which is sufficient to extract the scaling behavior established in 1D.

\end{document}